\documentclass[11pt]{article}
\usepackage[a4paper]{geometry}
\usepackage{amsfonts, amsmath, amssymb, amsthm, graphicx, caption, authblk, multirow, makecell, framed, float, xcolor, enumitem, tikz, hyperref}
\theoremstyle{definition}

\theoremstyle{remark}

\newlength{\ralen}
\newcommand{\crd}[1]{\raisebox{\ralen}{\framebox(12,13){#1}}}

\newcommand{\aaa}{\crd{$\alpha$}}
\newcommand{\bbb}{\crd{$\beta$}}
\newcommand{\back}{\crd{?}}
\newcommand{\backk}{\crd{?}}

\newcommand{\acard}[1]{\overset{#1}{\backbb}}
\newcommand{\cardbb}[2]{\overset{\crd{#1}}{\crd{#2}}}
\newcommand{\cardbbb}[3]{\overset{\overset{\crd{#1}}{\crd{#2}}}{\crd{#3}}}
\newcommand{\backbb}{\cardbb{?}{?}}
\newcommand{\backbbb}{\cardbbb{?}{?}{?}}

\begin{document}
\title{Card-Based Computation in the Virtual Player Simulation Model}
\author[1]{Suthee Ruangwises\thanks{\texttt{suthee@cp.eng.chula.ac.th}}}
\affil[1]{Department of Computer Engineering, Faculty of Engineering, Chulalongkorn University, Bangkok, Thailand}
\date{}
\maketitle

\begin{abstract}
Player simulation has recently emerged as a new direction in card-based cryptography, with protocols developed for simulating virtual players in physical card games such as Old Maid, UNO, and President. Unlike conventional card-based secure computation, player simulation imposes additional constraints: the cards represent a persistent game state, the remaining cards in a virtual player's hand must be preserved after each action, and it is desirable to represent each card in the game by a single physical card. In this paper, we study generic card-based computation in the virtual player simulation model. We focus on games whose cards admit a publicly known ranking and propose two fundamental protocols. First, we present the Play-Minimum protocol, which securely selects and plays the minimum-value card from a virtual player's hand when all cards in the deck have distinct values. By symmetry, the protocol can also be used to play the maximum-value card. Second, we present the Sorting protocol, which securely arranges a virtual player's hand in nondecreasing order and remains applicable when multiple cards have the same value. These protocols provide generic computational primitives independent of any particular card game and constitute a step toward understanding the computational capabilities of the virtual player simulation model.

\end{abstract}

\section{Introduction}
\emph{Card-based cryptography} studies cryptographic protocols that are implemented using a physical deck of playing cards. In such protocols, the faces of cards encode secret information, while physical operations such as shuffling, rearranging, and turning over cards are used to perform computation. A key advantage of card-based protocols is that they can be executed without computers or other electronic devices. Moreover, because every operation can be directly observed by the participants, the correctness of a protocol can be verified through simple physical actions rather than relying on the implementation of cryptographic software. These properties make card-based protocols suitable for situations where computation must be performed in a transparent and intuitive manner using only physical objects. Historically, research on card-based cryptography has developed along two primary directions: secure multi-party computation and zero-knowledge proofs. Recently, player simulation has emerged as a potential third direction, motivated by the problem of securely simulating virtual players in physical card games.

\textbf{Secure Multi-Party Computation.} One major direction of card-based cryptography concerns secure multi-party computation, where multiple parties jointly compute a function of their private inputs without revealing the inputs themselves. Protocols have been proposed for various Boolean functions, including logical AND \cite{mizuki16,mizuki09}, logical XOR \cite{mizuki09}, the \emph{majority function} \cite{nishida13,toyoda}, and the \emph{equality function} \cite{ruangwises21}. Beyond individual functions, Nishida et al.~\cite{nishida15} constructed a protocol that can compute any Boolean function of $n$ variables using $2n+6$ cards. Later, Shinagawa and Nuida~\cite{garbled} showed that any Boolean function can be computed using only a single shuffle.

\textbf{Zero-Knowledge Proof.} Another major direction concerns card-based zero-knowledge proofs, in which a prover demonstrates to a verifier that they know a solution to a given problem without revealing the solution itself \cite{zkp}. Such protocols have been proposed for computational problems such as graph isomorphism \cite{graph} and pancake sorting \cite{pancake}, as well as for pencil puzzles such as Sudoku \cite{ono,sudoku2,sudoku,tanaka} and Nonogram \cite{nonogram}. Card-based zero-knowledge proofs have also recently been extended to mobile puzzle games, such as Ball Sort Puzzle \cite{ball}.

\subsection{Player Simulation Protocol}
Many card games can be played with as few as two players, but become more engaging with a larger number of participants. When there are not enough human players, virtual players can be introduced to simulate additional participants. While such virtual players can be easily implemented by software in digital card games, realizing them in a purely physical setting without electronic devices is considerably more challenging. In particular, no human player should be allowed to inspect the virtual player's hand, as this would reveal information that is intended to remain secret. At the same time, the human players must be able to ensure that the virtual player follows the rules of the game correctly. A player simulation protocol must therefore enable the virtual player to perform valid actions while keeping its private information hidden from all human players.

From a theoretical perspective, a player simulation protocol can be viewed as a specialized form of secure multi-party computation in which a collection of face-down cards represents the private state of a virtual player. Based on this hidden state and publicly available information about the current game state, the participants jointly execute a prescribed sequence of card operations to determine an action for the virtual player, without entrusting its private information to any individual participant. The protocol should reveal only the information that would normally become public as a consequence of the action, while keeping the remaining cards in the virtual player's hand intact for subsequent turns. Moreover, although the distribution of cards among the players is private, the union of all cards in their hands plus the unplayed cards is typically public information, as it is determined by the deck and the cards that have already been revealed.

Another desirable, though not strictly necessary, requirement is a one-to-one correspondence between cards in the game and physical cards used by the protocol. This is in contrast to conventional card-based secure computation and zero-knowledge protocols, where multiple cards are often used to encode a single value. Such a restriction considerably limits the available operations, as many standard techniques, including those based on the commonly used two-card encoding of a binary value, cannot be directly applied. Thus, although player simulation can be regarded as a form of secure multi-party computation, its requirements impose a distinct set of constraints that motivates studying it as a computational model in its own right.

In 2025, Shinagawa et al.~\cite{oldmaid} introduced the first card-based protocol for simulating virtual players, focusing on the card game \emph{Old Maid}. Their protocol allows pairs of cards with the same number to be identified and removed from the virtual player's hand without revealing any additional information. Ruangwises and Shinagawa~\cite{uno} subsequently developed a player simulation protocol for \emph{UNO}, where a virtual player must choose a valid card according to the current game state and randomly select among multiple available choices. More recently, Miyahara et al.~\cite{president} proposed a player simulation protocol for the card game \emph{President}, extending the framework to support actions in which multiple cards are played simultaneously.

Existing player simulation protocols have primarily been designed for specific card games. Consequently, the computational capabilities of the player simulation model itself remain largely unexplored. In particular, even fundamental operations that naturally arise when determining a player's action such as finding a minimum or maximum value and sorting a collection of values have not yet been systematically studied in this model. Developing generic protocols for such operations would not only clarify the computational power of player simulation protocols but also provide reusable building blocks for constructing virtual players for a wider variety of physical games.

\subsection{Our Contribution}
In this paper, we study generic card-based computations in the player simulation model, focusing on the setting where all cards in the deck are rankable, i.e. there is a publicly known ordering among their values. We develop two fundamental protocols for manipulating a virtual player's hand while preserving its secrecy.

First, we propose the \emph{Play-Minimum protocol}, which identifies and plays a card of minimum rank from the virtual player's hand without revealing the remaining cards. This protocol applies when all cards in the deck are distinct. By reversing the ordering of the cards, the same protocol can symmetrically be used to play a card of maximum rank.

Second, we propose the \emph{Sorting protocol}, which rearranges all cards in the virtual player's hand in nondecreasing order of their ranks without revealing any card to the human players. Unlike the Play-Minimum protocol, the Sorting protocol also applies to decks containing multiple cards of the same rank.

These protocols provide generic building blocks for virtual player simulation that are independent of the rules of any particular card game. They also demonstrate that nontrivial computations on a hidden hand can be performed under the restrictive requirements of the player simulation model.

\section{Preliminaries}
\subsection{Cards}
We use physical cards whose backs are indistinguishable. The cards used in our protocols are the actual cards from the card game being simulated. Our protocols also require additional cards that serve as auxiliary labels and markers, thus multiple copies of the physical deck may be needed.

We assume that the cards are \emph{rankable}: there is a publicly known total order on their values. Thus, for any two cards $c$ and $c'$, their values can be compared whenever their fronts are visible. The Play-Minimum protocol in Section~\ref{playmin} additionally assumes that all game cards have distinct values, whereas the Sorting protocol in Section~\ref{sort} allows multiple cards to have the same value.

\subsection{Pile-Scramble Shuffle}
Given a matrix of cards, a \emph{pile-scramble shuffle} \cite{scramble} permutes its columns according to a uniformly random permutation unknown to all parties. For example, consider a pile-scramble shuffle applied to the following $2 \times 3$ matrix:
$$
\acard{1}\, \acard{2}\, \acard{3}\,.
$$
	
After the shuffle, the resulting sequence is one of
$$
\acard{1}\, \acard{2}\, \acard{3}\,,
\quad
\acard{1}\, \acard{3}\, \acard{2}\,,
\quad
\acard{2}\, \acard{1}\, \acard{3}\,,
\quad
\acard{2}\, \acard{3}\, \acard{1}\,,
\quad
\acard{3}\, \acard{1}\, \acard{2}\,,
\text{ or} \quad
\acard{3}\, \acard{2}\, \acard{1}\,,
$$
each occurring with probability $1/6$.

In practice, this shuffle can be implemented by placing cards in each column into a separate envelope and randomly permuting the envelopes on a table. Alternatively, the cards in each column may be bound together using a paper clip or a rubber band before the piles are randomly permuted.

\subsection{Pile-Shifting Shuffle}
Given a matrix of cards, a \emph{pile-shifting shuffle} \cite{polygon} permutes its columns according to a uniformly random cyclic shift unknown to all parties. For example, consider a pile-shifting shuffle applied to the following $2 \times 3$ matrix:
$$
\acard{1}\, \acard{2}\, \acard{3}\,.
$$
	
After the shuffle, the resulting sequence is one of
$$
\acard{1}\, \acard{2}\, \acard{3}\,,
\quad
\acard{2}\, \acard{3}\, \acard{1}\,,
\text{ or} \quad
\acard{3}\, \acard{1}\, \acard{2}\,,
$$
each occurring with probability $1/3$.

In practice, this shuffle can be implemented by placing cards in each column into a separate envelope, stacking the envelopes, and having the participants take turns performing \emph{Hindu cuts}, i.e. moving an arbitrary number of envelopes from the bottom of the stack to the top \cite{hindu}.

\section{Play-Minimum Protocol} \label{playmin}
Suppose that there are $n-1$ players in the game. Let $P_1$ be the virtual player to be simulated, and let $P_2,P_3,\ldots,P_{n-1}$ be the other players, each of whom may be either human or virtual. We regard the unplayed deck as the hand of a hypothetical player $P_n$. Like the players' hands, its contents remain hidden throughout the game.

Let \aaa, \bbb, \crd{$\theta_1$}, \crd{$\theta_2$}, $\ldots$, \crd{$\theta_n$} be distinct, pre-designated types of cards. The cards \crd{$\theta_1$},\ldots,\crd{$\theta_n$} serve as ownership labels, while \aaa\ and \bbb\ are used to distinguish the cards belonging to $P_1$ from all other cards.

The virtual player $P_1$ is operated jointly by the participants according to the following procedure.
 
\begin{enumerate}
	\item Arrange all players' hands and the unplayed deck face-down in a single horizontal sequence, treating the unplayed deck as the hand of $P_n$.
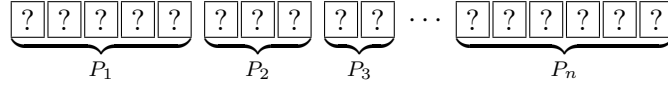
	$$\underbrace{\back\, \back\, \back\, \back\, \back}_{P_1}\, \, \underbrace{\back\, \back\, \back}_{P_2}\, \, \underbrace{\back\, \back}_{P_3}\, \, \cdots\, \, \underbrace{\back\, \back\, \back\, \back\, \back\, \back}_{P_n}$$
	
	\item For each $i \in \{1,2,\ldots,n\}$, place a card \crd{$\theta_i$} below every card belonging to $P_i$, forming the second row of the matrix.
	$$\cardbb{?}{$\theta_1$}\, \cardbb{?}{$\theta_1$}\, \cardbb{?}{$\theta_1$}\, \cardbb{?}{$\theta_1$}\, \cardbb{?}{$\theta_1$}\, \, \, \cardbb{?}{$\theta_2$}\, \cardbb{?}{$\theta_2$}\, \cardbb{?}{$\theta_2$}\, \, \, \cardbb{?}{$\theta_3$}\, \cardbb{?}{$\theta_3$}\, \, \cdots\, \, \cardbb{?}{$\theta_n$}\, \cardbb{?}{$\theta_n$}\, \cardbb{?}{$\theta_n$}\, \cardbb{?}{$\theta_n$}\, \cardbb{?}{$\theta_n$}\, \cardbb{?}{$\theta_n$}$$
	
	\item Place a card \aaa\ below each \crd{$\theta_1$}, and a card \bbb\ below each \crd{$\theta_i$} for $2 \leq i \leq n$, forming the third row of the matrix.
	$$\cardbbb{?}{$\theta_1$}{$\alpha$}\, \cardbbb{?}{$\theta_1$}{$\alpha$}\, \cardbbb{?}{$\theta_1$}{$\alpha$}\, \cardbbb{?}{$\theta_1$}{$\alpha$}\, \cardbbb{?}{$\theta_1$}{$\alpha$}\, \, \, \cardbbb{?}{$\theta_2$}{$\beta$}\, \cardbbb{?}{$\theta_2$}{$\beta$}\, \cardbbb{?}{$\theta_2$}{$\beta$}\, \, \, \cardbbb{?}{$\theta_3$}{$\beta$}\, \cardbbb{?}{$\theta_3$}{$\beta$}\, \, \cdots\, \, \cardbbb{?}{$\theta_n$}{$\beta$}\, \cardbbb{?}{$\theta_n$}{$\beta$}\, \cardbbb{?}{$\theta_n$}{$\beta$}\, \cardbbb{?}{$\theta_n$}{$\beta$}\, \cardbbb{?}{$\theta_n$}{$\beta$}\, \cardbbb{?}{$\theta_n$}{$\beta$}$$
	
	\item Turn all cards face-down.
	$$\backbbb\, \backbbb\, \backbbb\, \backbbb\, \backbbb\, \, \, \backbbb\, \backbbb\, \backbbb\, \, \, \backbbb\, \backbbb\, \, \cdots\, \, \backbbb\, \backbbb\, \backbbb\, \backbbb\, \backbbb\, \backbbb$$
	
	\item Apply a pile-scramble shuffle to the matrix.
	$$\left[\begin{tabular}{c|c|c|c|c|c|c|c|c}
		\ \backk \ \ & \ \backk \ \ & \ \backk \ \ & \ \backk \ \ & \ \backk \ \ & \ \backk \ \ & \ \backk \ \ & \ \ \ & \ \backk \ \ \\
		\ \backk \ \ & \ \backk \ \ & \ \backk \ \ & \ \backk \ \ & \ \backk \ \ & \ \backk \ \ & \ \backk \ \ & \ \ \ & \ \backk \ \ \\
		\ \backk \ \ & \ \backk \ \ & \ \backk \ \ & \ \backk \ \ & \ \backk \ \ & \ \backk \ \ & \ \backk \ \ & \ $\cdots$ \ \ & \ \backk \ \
	\end{tabular}\right]$$
    	
	\item Turn over all cards in the first row, revealing their identities while keeping their ownership hidden.
	$$\cardbbb{8}{?}{?}\, \, \, \, \, \cardbbb{9}{?}{?}\, \, \, \, \, \cardbbb{2}{?}{?}\, \, \, \, \, \cardbbb{5}{?}{?}\, \, \, \, \, \cardbbb{1}{?}{?}\, \, \, \, \, \cardbbb{$m$}{?}{?}\, \, \, \, \, \cardbbb{4}{?}{?}\, \, \, \, \, \cdots\, \, \, \, \, \cardbbb{7}{?}{?}$$
	
	\item Sort the columns in increasing order according to the values of the cards in the first row.
	$$\cardbbb{1}{?}{?}\, \, \, \, \, \cardbbb{2}{?}{?}\, \, \, \, \, \cardbbb{4}{?}{?}\, \, \, \, \, \cardbbb{5}{?}{?}\, \, \, \, \, \cardbbb{7}{?}{?}\, \, \, \, \, \cardbbb{8}{?}{?}\, \, \, \, \, \cardbbb{9}{?}{?}\, \, \, \, \, \cdots\, \, \, \, \, \cardbbb{$m$}{?}{?}$$
	
	\item Starting from the first column, turn over the card in the third row. If it is an \aaa, stop and play the card in the first row of that column. Otherwise, proceed to the next column and repeat until the first \aaa\ is found.
	$$\cardbbb{1}{?}{$\beta$}\, \, \, \, \, \cardbbb{2}{?}{$\beta$}\, \, \, \, \, \cardbbb{4}{?}{$\beta$}\, \, \, \, \, \cardbbb{5}{?}{$\alpha$}\, \, \, \, \, \cardbbb{7}{?}{?}\, \, \, \, \, \cardbbb{8}{?}{?}\, \, \, \, \, \cardbbb{9}{?}{?}\, \, \, \, \, \cdots\, \, \, \, \, \cardbbb{$m$}{?}{?}$$
	
	\item Remove the column containing the first \aaa, and turn all cards face-down.
	$$\cardbbb{?}{?}{?}\, \, \, \, \, \cardbbb{?}{?}{?}\, \, \, \, \, \cardbbb{?}{?}{?}\, \, \, \, \, \cardbbb{?}{?}{?}\, \, \, \, \, \cardbbb{?}{?}{?}\, \, \, \, \, \cardbbb{?}{?}{?}\, \, \, \, \, \cdots\, \, \, \, \, \cardbbb{?}{?}{?}$$
	
	\item Apply another pile-scramble shuffle to the matrix.
	$$\left[\begin{tabular}{c|c|c|c|c|c|c|c}
		\ \backk \ \ & \ \backk \ \ & \ \backk \ \ & \ \backk \ \ & \ \backk \ \ & \ \backk \ \ & \ \ \ & \ \backk \ \ \\
		\ \backk \ \ & \ \backk \ \ & \ \backk \ \ & \ \backk \ \ & \ \backk \ \ & \ \backk \ \ & \ \ \ & \ \backk \ \ \\
		\ \backk \ \ & \ \backk \ \ & \ \backk \ \ & \ \backk \ \ & \ \backk \ \ & \ \backk \ \ & \ $\cdots$ \ \ & \ \backk \ \
	\end{tabular}\right]$$
	
	\item Turn over all cards in the second row, revealing the ownership label $\theta_i$ of each column. Return each card in the first row to the hand of the player indicated by its corresponding label.
	$$\cardbbb{?}{$\theta_2$}{?}\, \, \, \, \, \cardbbb{?}{$\theta_1$}{?}\, \, \, \, \, \cardbbb{?}{$\theta_4$}{?}\, \, \, \, \, \cardbbb{?}{$\theta_1$}{?}\, \, \, \, \, \cardbbb{?}{$\theta_1$}{?}\, \, \, \, \, \cardbbb{?}{$\theta_3$}{?}\, \, \, \, \, \cdots\, \, \, \, \, \cardbbb{?}{$\theta_2$}{?}$$
\end{enumerate}

Let $k$ denote the total number of cards remaining in the game. The protocol requires $2k$ additional cards and uses two pile-scramble shuffles.

Note that Step~6 reveals the set of cards remaining in the players' hands and the unplayed deck. This information is already determined by the initial deck and the cards that have previously been discarded, all of which are public information. Hence, Step~6 reveals no information beyond what can in principle already be deduced from the public game history.

In practice, however, human players may not remember all previously played cards, and explicitly revealing the remaining cards may therefore provide information that they would not otherwise readily possess. To preserve this practical uncertainty and provide a more natural gameplay experience, the discard pile can instead be treated as the hand of an additional hypothetical player $P_{n+1}$. In this variant, a card with $\theta_{n+1}$ label is placed below each discarded card in Step~2, so that the first row contains the entire original deck. Consequently, revealing and sorting the first row discloses no information about which cards remain in play.

By symmetry, the Play-Minimum protocol can also be used to play the maximum-value card in $P_1$'s hand. It suffices to reverse the ordering in Step~7, sorting the columns from maximum to minimum. The first \aaa\ encountered in Step~8 then corresponds to the maximum-value card in $P_1$'s hand. We refer to this variant as the \emph{Play-Maximum protocol}.

\subsection{Proof of Correctness and Security}
We first prove the correctness of the Play-Minimum protocol. By the construction in Steps~2 and~3, for each card $c$ in the first row, the card \crd{$\theta_i$} in the second row indicates that $c$ belongs to $P_i$, while the card in the third row indicates whether $c$ belongs to $P_1$: an \aaa\ if it does and a \bbb\ otherwise. In Step~7, the columns are sorted in increasing order according to the values of the cards in the first row. Therefore, the first \aaa\ encountered in Step~8 corresponds precisely to the minimum-value card in $P_1$'s hand, which is then played. Moreover, in Step~11, each remaining card in the first row is returned to the player indicated by the corresponding \crd{$\theta_i$} in the second row. Thus, all other cards are returned to their original owners, and the protocol is correct.

The Play-Minimum protocol follows the computational model of card-based cryptography \cite{formal}, in which security is defined information-theoretically. It therefore suffices to show that the cards turned face-up during the execution reveal no information beyond what is already public or implied by the output of the protocol.

\begin{itemize}
\item In Step~6, all cards in the first row, i.e. all cards remaining in the game, are revealed. Due to the pile-scramble shuffle in Step~5, the $k$ columns are uniformly distributed over all $k!$ permutations. Hence, the revealed order contains no information about the ownership of the cards. The only information revealed is the set of remaining cards, which is public information since it is determined by the full deck and the public discard pile.

\item In Step~8, the cards in the third row are revealed sequentially until the first \aaa\ is encountered. Suppose that the card played in this step has value $x$. Since all cards in the deck have distinct values, every card preceding $x$ in the sorted sequence has value strictly smaller than $x$. The revealed \bbb s therefore indicate only that none of these smaller cards belongs to $P_1$. This information is already implied by the fact that $x$ is the minimum-value card in $P_1$'s hand. Thus, Step~8 reveals no information beyond the card played by $P_1$.

\item In Step~11, all cards in the second row are revealed. Due to the pile-scramble shuffle in Step~10, the remaining columns are uniformly distributed over all permutations. Hence, the revealed arrangement contains no information about which particular cards belong to which players. The only information revealed is the number of cards held by each player, which is public information.
\end{itemize}

Therefore, the Play-Minimum protocol is both correct and secure.

\section{Sorting Protocol} \label{sort}
The Play-Minimum protocol from Section~\ref{playmin} relies on the assumption that all cards in the deck have distinct values. When multiple cards may have the same value, the protocol can reveal additional information about the virtual player's hand. For example, suppose that there are three cards of value $1$, which is the minimum value in the deck. If, after sorting, the third-row cards corresponding to these three cards are revealed as \bbb, \bbb, and \aaa, respectively, then the players learn that $P_1$ holds exactly one card of value $1$, namely the card being played. This information is not implied by the played card alone and therefore constitutes an information leak.

Moreover, playing a minimum- or maximum-value card is not sufficient for every possible strategy. A virtual player may need to select a card according to its relative position within the hand, such as playing a median-value card. To support such strategies, we develop a \emph{Sorting protocol} that securely arranges the cards in the virtual player's hand in nondecreasing order. Unlike the Play-Minimum protocol, the Sorting protocol remains applicable when multiple cards in the deck have the same value.

Suppose that there are $n-1$ players in the game. Let $P_1$ be the virtual player to be simulated, and let $P_2,P_3,\ldots,P_{n-1}$ be the other players. Again, we regard the unplayed deck as the hand of a hypothetical player $P_n$.

Let \aaa, \crd{$\theta_1$}, \crd{$\theta_2$}, $\ldots$, \crd{$\theta_n$} be distinct, pre-designated types of cards. The cards \crd{$\theta_1$},\ldots,\crd{$\theta_n$} serve as ownership labels, while \aaa\ is used as a marker for restoring the cyclic order of the matrix.

The virtual player $P_1$ is operated jointly by the participants according to the following procedure.
 
\begin{enumerate}
	\item Arrange all players' hands and the unplayed deck face-down in a single horizontal sequence, treating the unplayed deck as the hand of $P_n$.
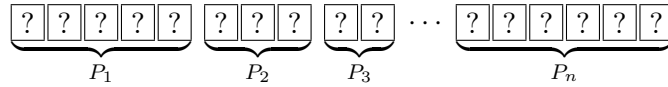
	$$\underbrace{\back\, \back\, \back\, \back\, \back}_{P_1}\, \, \underbrace{\back\, \back\, \back}_{P_2}\, \, \underbrace{\back\, \back}_{P_3}\, \, \cdots\, \, \underbrace{\back\, \back\, \back\, \back\, \back\, \back}_{P_n}$$
	
	\item For each $i \in \{1,2,\ldots,n\}$, place a card \crd{$\theta_i$} below every card belonging to $P_i$, forming the second row of the matrix.
	$$\cardbb{?}{$\theta_1$}\, \cardbb{?}{$\theta_1$}\, \cardbb{?}{$\theta_1$}\, \cardbb{?}{$\theta_1$}\, \cardbb{?}{$\theta_1$}\, \, \, \cardbb{?}{$\theta_2$}\, \cardbb{?}{$\theta_2$}\, \cardbb{?}{$\theta_2$}\, \, \, \cardbb{?}{$\theta_3$}\, \cardbb{?}{$\theta_3$}\, \, \cdots\, \, \cardbb{?}{$\theta_n$}\, \cardbb{?}{$\theta_n$}\, \cardbb{?}{$\theta_n$}\, \cardbb{?}{$\theta_n$}\, \cardbb{?}{$\theta_n$}\, \cardbb{?}{$\theta_n$}$$
	
	\item Turn all cards face-down.
	$$\backbb\, \backbb\, \backbb\, \backbb\, \backbb\, \, \, \backbb\, \backbb\, \backbb\, \, \, \backbb\, \backbb\, \, \cdots\, \, \backbb\, \backbb\, \backbb\, \backbb\, \backbb\, \backbb$$
	
	\item Apply a pile-scramble shuffle to the matrix.
	$$\left[\begin{tabular}{c|c|c|c|c|c|c|c|c}
		\ \backk \ \ & \ \backk \ \ & \ \backk \ \ & \ \backk \ \ & \ \backk \ \ & \ \backk \ \ & \ \backk \ \ & \ \ \ & \ \backk \ \ \\
		\ \backk \ \ & \ \backk \ \ & \ \backk \ \ & \ \backk \ \ & \ \backk \ \ & \ \backk \ \ & \ \backk \ \ & \ $\cdots$ \ \ & \ \backk \ \
	\end{tabular}\right]$$
    	
	\item Turn over all cards in the first row, revealing their identities while keeping their ownership hidden.
	$$\cardbb{4}{?}\, \, \, \, \, \cardbb{3}{?}\, \, \, \, \, \cardbb{8}{?}\, \, \, \, \, \cardbb{1}{?}\, \, \, \, \, \cardbb{4}{?}\, \, \, \, \, \cardbb{6}{?}\, \, \, \, \, \cardbb{1}{?}\, \, \, \, \, \cdots\, \, \, \, \, \cardbb{4}{?}$$
	
	\item Sort the columns in nondecreasing order according to the values of the cards in the first row. Columns whose first-row cards have the same value may be arranged in an arbitrary order.
	$$\cardbb{1}{?}\, \, \, \, \, \cardbb{1}{?}\, \, \, \, \, \cardbb{3}{?}\, \, \, \, \, \cardbb{4}{?}\, \, \, \, \, \cardbb{4}{?}\, \, \, \, \, \cardbb{4}{?}\, \, \, \, \, \cardbb{6}{?}\, \, \, \, \, \cdots\, \, \, \, \, \cardbb{$m$}{?}$$
	
	\item Append one \aaa\ to the right end of each row, forming an additional rightmost column.
	$$\cardbb{1}{?}\, \, \, \, \, \cardbb{1}{?}\, \, \, \, \, \cardbb{3}{?}\, \, \, \, \, \cardbb{4}{?}\, \, \, \, \, \cardbb{4}{?}\, \, \, \, \, \cardbb{4}{?}\, \, \, \, \, \cardbb{6}{?}\, \, \, \, \, \cdots\, \, \, \, \, \cardbb{$m$}{?}\, \, \, \, \, \cardbb{$\alpha$}{$\alpha$}$$
	
	\item Turn all cards face-down, and apply a pile-shifting shuffle to the matrix.
	$$\left\langle\begin{tabular}{c|c|c|c|c|c|c|c|c}
		\ \backk \ \ & \ \backk \ \ & \ \backk \ \ & \ \backk \ \ & \ \backk \ \ & \ \backk \ \ & \ \backk \ \ & \ \ \ & \ \backk \ \ \\
		\ \backk \ \ & \ \backk \ \ & \ \backk \ \ & \ \backk \ \ & \ \backk \ \ & \ \backk \ \ & \ \backk \ \ & \ $\cdots$ \ \ & \ \backk \ \
	\end{tabular}\right\rangle$$
	
	\item Turn over the card in the second row of the rightmost column. If it is a \crd{$\theta_i$} for some $i \geq 2$, return the card above it to $P_i$ and remove the rightmost column from the matrix. If it is a \crd{$\theta_1$} or an \aaa, leave the column unchanged. Then, return to Step~8.
	$$\cardbb{?}{?}\, \, \, \, \, \cardbb{?}{?}\, \, \, \, \, \cardbb{?}{?}\, \, \, \, \, \cardbb{?}{?}\, \, \, \, \, \cardbb{?}{?}\, \, \, \, \, \cardbb{?}{?}\, \, \, \, \, \cardbb{?}{?}\, \, \, \, \, \cdots\, \, \, \, \, \cardbb{?}{$\theta_2$}$$
	
	\item Repeat Steps~8--9 until exactly $k_1+1$ columns remain in the matrix, where $k_1$ is the number of cards in $P_1$'s hand.
	
	\item Turn over all cards in the second row. Cyclically shift the columns so that the unique \aaa\ in the second row is in the rightmost column.
	$$\cardbb{?}{$\theta_1$}\, \, \, \, \, \cardbb{?}{$\theta_1$}\, \, \, \, \, \cardbb{?}{$\theta_1$}\, \, \, \, \, \cardbb{?}{$\theta_1$}\, \, \, \, \, \cardbb{?}{$\theta_1$}\, \, \, \, \, \cardbb{?}{$\alpha$}$$
	
	\item Remove the rightmost column. The first row now consists precisely of the cards in $P_1$'s hand, arranged in nondecreasing order.
	$$\cardbb{?}{$\theta_1$}\, \, \, \, \, \cardbb{?}{$\theta_1$}\, \, \, \, \, \cardbb{?}{$\theta_1$}\, \, \, \, \, \cardbb{?}{$\theta_1$}\, \, \, \, \, \cardbb{?}{$\theta_1$}$$
\end{enumerate}

Let $k$ be the total number of cards remaining in the game, and let $k_1$ be the number of cards in $P_1$'s hand. The protocol requires $k+2$ additional cards. It uses one pile-scramble shuffle in Step~4 and a variable number of pile-shifting shuffles in Step~8. Suppose that $i$ columns not belonging to $P_1$ have already been removed, where $0\leq i<k-k_1$. At this point, $k+1-i$ columns remain, of which $k-k_1-i$ can be removed in Step~9. Hence, the probability that one execution of Step~8 leads to a removal is
$$\frac{k-k_1-i}{k+1-i},$$
and the expected number of pile-shifting shuffles required for the next removal is $(k+1-i)/(k-k_1-i)$. Therefore, the expected total number of pile-shifting shuffles is

\begin{align*}
\sum_{i=0}^{k-k_1-1}\frac{k+1-i}{k-k_1-i} &= (k-k_1)+\sum_{i=0}^{k-k_1-1}\frac{k_1+1}{k-k_1-i} \\
&= (k-k_1)+(k_1+1)\sum_{j=1}^{k-k_1}\frac{1}{j}.
\end{align*}

Since $\sum_{j=1}^{k-k_1} 1/j = O(\log (k-k_1))$, the expected number of pile-shifting shuffles is $O(k\log k)$.

The additional \aaa\ column introduced in Step~7 serves as a reference point for restoring the linear order of the cards. Each pile-shifting shuffle cyclically shifts the columns and therefore preserves their cyclic order. Consequently, as columns belonging to players other than $P_1$ are removed, the relative cyclic order of the remaining columns is preserved. Once only the cards of $P_1$ and the \aaa\ column remain, the latter identifies the boundary of the sorted sequence, allowing the cards of $P_1$ to be restored to their nondecreasing linear order in Step~11.

As in the Play-Minimum protocol, Step~5 reveals the cards remaining in the players' hands and the unplayed deck. Since the discard pile is public, this set is determined by the original deck and the previously discarded cards and is therefore public information. Nevertheless, if one wishes to avoid explicitly revealing information that human players may not have kept track of during gameplay, the discard pile may again be treated as the hand of an additional hypothetical player $P_{n+1}$. By including the discarded cards in the matrix and labeling them with $\theta_{n+1}$, the first row contains the entire original deck, so revealing it provides no additional information about which cards remain in play.

\subsection{Proof of Correctness and Security}
We first prove the correctness of the Sorting protocol. By the construction in Step~2, for each card $c$ in the first row, the card \crd{$\theta_i$} in the second row indicates that $c$ belongs to $P_i$. In Step~6, all columns are sorted in nondecreasing order according to the values of the cards in the first row. The pile-shifting shuffles performed subsequently preserve the cyclic order of the columns. Moreover, removing a column in Step~9 does not change the relative cyclic order of the remaining columns. Therefore, after all cards not belonging to $P_1$ have been removed, the cards belonging to $P_1$ remain in nondecreasing cyclic order together with the \aaa\ column. In Step~11, the columns are cyclically shifted so that the \aaa\ column is moved to the rightmost position, thereby restoring the original linear order established in Step~6. After removing the \aaa\ column in Step~12, the first row consists precisely of the cards in $P_1$'s hand arranged in nondecreasing order. Hence, the protocol is correct.

We next prove the security of the protocol. As in the Play-Minimum protocol, the Sorting protocol follows the computational model of card-based cryptography \cite{formal}, in which security is defined information-theoretically. It therefore suffices to show that the cards turned face-up during the execution reveal no information beyond what is already public.

\begin{itemize}
\item In Step~5, all cards in the first row, i.e. all cards remaining in the game, are revealed. Due to the pile-scramble shuffle in Step~4, the $k$ columns are uniformly distributed over all $k!$ permutations. Hence, the revealed order contains no information about the ownership of the cards. The only information revealed is the multiset of remaining cards, which is public information since it is determined by the full deck and the public discard pile.

\item In each execution of Step~9, only the card in the second row of the rightmost column is revealed. Due to the pile-shifting shuffle in Step~8, every remaining column is equally likely to become the rightmost column. Therefore, the probability of revealing each ownership label is distributed solely according to the publicly known numbers of remaining cards belonging to each player. In particular, revealing a \crd{$\theta_i$} does not reveal the identity or value of the card above it. If $i\geq 2$, that column is immediately removed and its first-row card is returned privately to $P_i$. If the revealed card is \crd{$\theta_1$} or \aaa, the column remains in the matrix. Thus, the information revealed in Step~9 depends only on the publicly known hand sizes and not on the values or ordering of the hidden cards.

\item In Step~11, all remaining cards in the second row are revealed. By the stopping condition in Step~10, exactly $k_1+1$ columns remain. Since every column labeled \crd{$\theta_i$} for $i\geq2$ is eventually removed, these columns consist precisely of the $k_1$ columns labeled \crd{$\theta_1$} and the unique \aaa\ column. Also, due to the pile-shifting shuffle in Step~8, the \aaa\ column has an equal probability to be at each of the $k_1+1$ columns. Therefore, revealing the second row at this point provides no additional information.
\end{itemize}

Therefore, the protocol is both correct and secure.

\section{Future Work}
We studied generic card-based computation in the player simulation model. We proposed the Play-Minimum protocol, which securely plays the minimum-value card from a virtual player's hand, and the Sorting protocol, which securely sorts the cards in the hand and also supports repeated values.

For future work, it would be interesting to develop protocols for other fundamental operations, such as selecting the $k$-th smallest card without sorting the entire hand and counting the number of cards with a particular property. Another direction is to improve the efficiency of our protocols in terms of the number of additional cards and shuffles. More broadly, characterizing which functions can be securely computed under the restrictions of the virtual player simulation model remains an important open problem.

\subsubsection*{Acknowledgement}
This work was supported by the 111th Anniversary Engineering Research Catalyst Fund Towards U Top 100.

\end{document}